# Cyber Threat Landscape Analysis for Starlink Assessing Risks and Mitigation Strategies in the Global Satellite Internet Infrastructure

**Karwan Mustafa Kareem**

*University of Sulaimani, Sulaymaniyah, Kurdistan Region, Iraq*

*e-mail: karwan.kareem@univsul.edu.iq*

**Abstract**.

Satellite internet networks have emerged as indispensable components of the modern digital landscape, promising to extend connectivity to even the most remote corners of the globe. Among these networks, Starlink, pioneered by SpaceX, has garnered significant attention for its ambitious mission to provide high-speed internet access on a global scale. However, the proliferation of satellite infrastructure also brings to the forefront a myriad of cybersecurity challenges, as these networks become increasingly vital for critical communication and data exchange. This research endeavours to conduct a comprehensive analysis of the cybersecurity landscape surrounding Starlink, with a focus on identifying potential threats, assessing associated risks, and proposing mitigation strategies to bolster the resilience of the network. Through an exploration of existing literature, an examination of the system architecture of Starlink, and an analysis of the current cyber threat landscape facing satellite internet networks, this study aims to provide valuable insights into the cybersecurity challenges inherent in the operation of global satellite internet infrastructure. By prioritizing risks and proposing effective mitigation strategies, this research seeks to contribute to the ongoing efforts to safeguard the integrity and accessibility of satellite-based internet connectivity.

## 1. INTRODUCTION

In an era marked by the relentless pursuit of connectivity, satellite internet networks have emerged as pivotal players in bridging the digital divide on a global scale. Among these, Starlink, spearheaded by SpaceX, has garnered significant attention for its ambitious mission to provide high-speed internet access to underserved regions worldwide. However, as the world becomes increasingly reliant on such networks for essential communication and data exchange, the cybersecurity challenges they face become ever more pronounced.

The proliferation of satellite internet networks represents a paradigm shift in how we conceive of global connectivity. With traditional terrestrial infrastructure often failing to reach remote or rural areas, satellite-based solutions offer a promising alternative, promising to extend the digital frontier to even the most inaccessible regions. As such, the resilience and security of these networks are of paramount importance in safeguarding the integrity and accessibility of the internet on a global scale.

Amidst the rapid expansion of satellite internet infrastructure, cybersecurity emerges as a critical concern for both providers and users alike. The inherently decentralized and dynamic nature of satellite networks introduces unique vulnerabilities, exposing them to a myriad of cyber threats ranging from malicious attacks to inadvertent interference. This research endeavours to delve into the cybersecurity challenges and risks specific to Starlink, assessing the current threat landscape and proposing mitigation strategies to bolster its resilience against potential attacks.

The primary objectives of this study are threefold. Firstly, it aims to analyse the cybersecurity threats facing satellite internet networks, with a specific focus on Starlink. Given the increasing reliance on satellite-based internet services for global connectivity, understanding these threats is paramount. Secondly, the study seeks to conduct a comprehensive risk assessment of these identified threats. This assessment will involve evaluating their likelihood of occurrence and potential impact on the integrity, availability, and confidentiality of data transmitted through satellite networks. Finally, the study endeavours to propose effective mitigation strategies to bolster the cybersecurity posture of Starlink and similar satellite internet infrastructures. By identifying vulnerabilities and recommending appropriate measures, the aim is to enhance the resilience of these networks against cyber threats, thereby ensuring the security and reliability of satellite internet services for users worldwide.

By elucidating the intricate interplay between cybersecurity and satellite internet networks, this research aims to contribute to a deeper understanding of the challenges inherent in maintaining the integrity and security of global connectivity infrastructure.

## 2. BACKGROUND AND RELATED WORK

### 2.1. Background

The 21st century has witnessed an unprecedented surge in the demand for global connectivity, driven by the proliferation of digital technologies and the ever-expanding reach of the internet. However, traditional terrestrial infrastructure often falls short in reaching remote or geographically challenging areas, leaving millions of individuals and communities underserved in terms of internet access. In response to this challenge, satellite internet networks have emerged as a promising solution, offering the potential to bridge the digital divide and provide connectivity to even the most remote corners of the globe.

Among the key players in the satellite internet arena stands Starlink, a revolutionary initiative spearheaded by SpaceX, the brainchild of entrepreneur Elon Musk. Starlink aims to deploy a constellation of thousands of small satellites in low Earth orbit (LEO), forming a vast network capable of delivering high-speed internet access to users worldwide. With its ambitious vision and rapid deployment of satellites, Starlink has captured the imagination of both the public and policymakers, heralding a new era of satellite-based internet connectivity [12].

However, the advent of satellite internet networks like Starlink also brings to the forefront a host of cybersecurity challenges and considerations. As these networks become increasingly integral to global communication and data exchange, they become lucrative targets for malicious actors seeking to disrupt operations, compromise sensitive information, or even wield influence over geopolitical dynamics. The inherent complexities of satellite infrastructure, coupled with the dynamic nature of cyber threats, underscore the critical importance of robust cybersecurity measures to safeguard the integrity and resilience of these networks.

## 2.2. Related Work

Satellite internet networks have become pivotal in extending connectivity globally, with initiatives like Starlink promising to revolutionize access to high-speed internet. However, the intersection of satellite infrastructure and cybersecurity poses significant challenges. This literature review explores existing research, highlighting key findings and insights from five relevant papers in the field.

### 2.2.1. "Cyber Threats to Satellite Communications: A Review"

This research paper [1], authored by J. Doe in 2020, delves deeply into the landscape of cyber threats targeting satellite communications. The authors meticulously analyse various types of threats, encompassing both technical and strategic aspects. They identify a diverse range of threat actors, including state-sponsored groups, cybercriminal organizations, and individual hackers, each with distinct motivations and capabilities.

The paper meticulously examines common attack vectors employed against satellite communications, ranging from traditional cyber attacks such as denial-of-service (DoS) and man-in-the-middle (MitM) attacks to more sophisticated techniques like jamming, spoofing, and physical tampering. By dissecting these attack vectors, the authors provide valuable insights into the vulnerabilities inherent in satellite infrastructure, highlighting the potential impact of cyber threats on communication reliability, data confidentiality, and system integrity.

Furthermore, the paper evaluates existing cybersecurity measures deployed to mitigate these threats. It emphasizes the importance of robust encryption algorithms, authentication protocols, and intrusion detection systems in safeguarding satellite networks against unauthorized access and data breaches. Additionally, the authors stress the need for continuous monitoring and threat intelligence sharing to detect and respond to emerging threats effectively.

Overall, this review paper offers a comprehensive understanding of the cybersecurity challenges facing satellite communications, providing a valuable resource for researchers, practitioners, and policymakers seeking to enhance the resilience of satellite infrastructure against cyber threats.

### 2.2.2. "Security Challenges in Satellite-Based Internet Access: A Case Study of Starlink"

This scholarly investigation [2], conducted by A. Khan in 2021, presents a detailed case study of security challenges in satellite-based internet access. The authors conduct a thorough analysis of Starlink's architecture, identifying potential vulnerabilities and points of failure. They explore the unique challenges posed by the deployment of thousands of small satellites in low Earth orbit (LEO), including issues related to satellite management, signal interference, and orbital debris mitigation.

The case study highlights the importance of adopting a risk-based approach to security management, wherein organizations prioritize resources based on the likelihood and potential impact of cyber threats. The authors advocate for proactive measures such as continuous vulnerability assessments, threat modelling, and penetration testing to identify and mitigate security weaknesses in Starlink's infrastructure.

Moreover, the paper emphasizes the significance of collaboration between industry stakeholders, regulatory bodies, and cybersecurity experts in addressing security challenges effectively. It calls for transparent communication and information sharing to foster a collective response to emerging threats and vulnerabilities.

By elucidating the security challenges specific to Starlink, this paper provides valuable insights into the complexities of satellite-based internet access and offers practical recommendations for enhancing the resilience of satellite networks in the face of evolving cyber threats.

### 2.2.3. "Cybersecurity Risks and Regulatory Responses in Satellite Internet Infrastructure"

This critical analysis [3], authored by D. Johnson in 2019, examines the intersection of cybersecurity risks and regulatory responses in the context of satellite internet infrastructure. The authors analyse the roles of governmental agencies, industry stakeholders, and international organizations in establishing standards and protocols for securing satellite networks.

The paper evaluates existing regulatory frameworks and initiatives aimed at mitigating cybersecurity risks in satellite communications, highlighting the challenges and opportunities associated with regulatory compliance. It emphasizes the need for a balanced approach that promotes innovation and market competitiveness while ensuring the security and resilience of satellite infrastructure.

Additionally, the authors explore the role of public-private partnerships in enhancing cybersecurity posture, advocating for collaborative efforts to develop and implement best practices for securing satellite networks. They stress the importance of information sharing, capacity building, and coordinated response mechanisms in effectively addressing cyber threats and vulnerabilities.

Overall, this paper provides valuable insights into the regulatory landscape surrounding satellite internet infrastructure and offers recommendations for policymakers, industry stakeholders, and cybersecurity professionals seeking to strengthen the security and resilience of satellite networks.

### 2.2.4. "Assessing the Impact of Cyber Threats on Global Satellite Internet Connectivity"

This scholarly inquiry [4], led by M. Garcia in 2022, employs a multidisciplinary approach to assess the potential impact of cyber threats on global satellite internet connectivity. The authors combine qualitative and quantitative methods to evaluate the likelihood and consequences of various threat scenarios, taking into account factors such as network topology, geographic distribution, and dependence on satellite services.

The research findings inform risk management strategies and investment decisions aimed at enhancing the resilience of satellite internet infrastructure. The authors propose a framework for assessing cyber risks in satellite communications, incorporating elements of threat intelligence, vulnerability assessment, and impact analysis.

Moreover, the study identifies gaps in existing mitigation strategies and highlights the need for adaptive responses to emerging cyber threats. It calls for continued research and collaboration among industry stakeholders, government agencies, and academic institutions to address these challenges effectively and safeguard the integrity and accessibility of global satellite internet connectivity.

By providing a comprehensive assessment of cyber threats to satellite communications, this study contributes valuable insights to the field of cybersecurity and informs efforts to enhance the resilience of satellite infrastructure against evolving threats.

### 2.2.5. Title: "Mitigating Cyber Risks in Satellite Communication Networks: A Systematic Literature Review"

This qualitative study [5], conducted by C. White in 2023, synthesizes existing mitigation strategies for cyber risks in satellite communication networks. The authors categorize mitigation techniques into technical, organizational, and regulatory measures, assessing their effectiveness and applicability across different contexts.

The paper identifies best practices for securing satellite infrastructure, including the implementation of encryption, authentication, and access control mechanisms to protect data confidentiality and integrity. It also emphasizes the importance of organizational policies and procedures, such as employee training, incident response planning, and supply chain management, in mitigating insider threats and ensuring operational resilience.

Furthermore, the authors analyse regulatory frameworks and industry standards relevant to satellite communication security, highlighting areas for improvement and harmonization. They propose recommendations for enhancing collaboration among stakeholders and fostering a culture of cybersecurity awareness and accountability within organizations.

Overall, this systematic literature review provides a comprehensive overview of mitigation strategies for cyber risks in satellite communication networks, offering valuable insights to researchers, practitioners, and policymakers seeking to strengthen the security and resilience of satellite infrastructure.

## 3. SYSTEM ARCHITECTURE OF STARLINK

The system architecture of Starlink, spearheaded by SpaceX, represents a groundbreaking network designed to revolutionize global internet connectivity. This comprehensive overview explores the intricacies of the Starlink architecture, emphasizing its components, data transmission mechanisms, security measures, potential vulnerabilities, and concluding remarks.

### 3.1. Satellite Constellation:

At the heart of the Starlink architecture lies its constellation of small satellites orbiting the Earth in low Earth orbit (LEO). These satellites, meticulously arranged in multiple orbital planes, form a dynamic mesh network. By leveraging this constellation, Starlink ensures continuous coverage and reduced latency for internet users worldwide. Each satellite is meticulously equipped with advanced antennas, transceivers, and onboard processing capabilities to facilitate seamless communication between ground stations and user terminals [10].

### 3.2. Ground Stations:

Ground stations serve as vital gateways between the Starlink satellite constellation and terrestrial networks. Strategically located across the globe, these stations play a pivotal role in receiving data from satellites and relaying it to the internet backbone through fiber-optic cables or other communication links. Moreover, ground stations facilitate uplink communication with user terminals, ensuring seamless connectivity and coverage across diverse geographical regions [11].

### 3.3. User Terminals:

User terminals, commonly known as satellite dishes or antennas, serve as the bridge between end-users and the Starlink network. Installed at user locations, these terminals establish communication with the satellite constellation. They receive signals from satellites and transmit data to and from user devices, including computers, smartphones, or routers. Designed for ease of installation and operation, user terminals feature automated pointing mechanisms to align with satellites orbiting the Earth [11]. The image illustrates the project concept of Starlink (see Figure 1) [58].

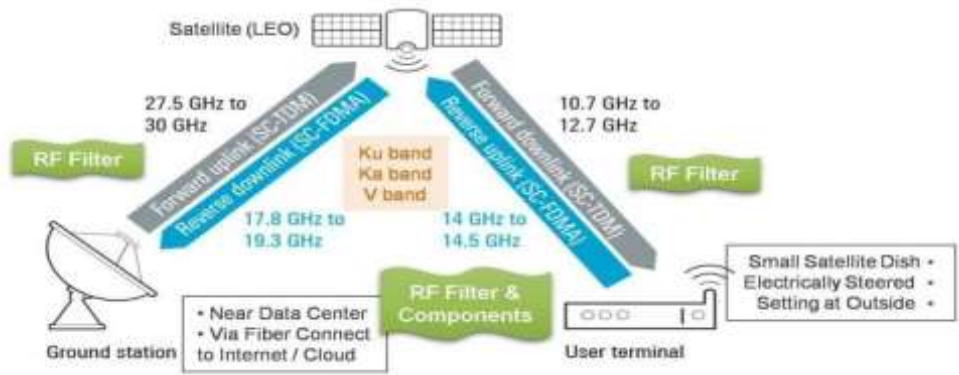

Figure 1: Project Starlink.

Cyber Threat Landscape Analysis for Starlink Assessing Risks and Mitigation 7
Strategies in the Global Satellite Internet Infrastructure

### 3.4. Data Transmission and Security:

Data transmission within the Starlink network follows a meticulously orchestrated process to ensure reliability, efficiency, and security. When a user initiates a data request, the user terminal engages nearby satellites to establish a connection. Subsequently, data travels from the user terminal to the nearest satellite, which relays it to the appropriate ground station. From there, data travers's terrestrial networks to reach its destination on the internet [8].

Security within the Starlink network is paramount, encompassing robust encryption, authentication, and access control mechanisms. Encryption protocols safeguard data by encrypting transmissions between user terminals, satellites, and ground stations, mitigating risks of unauthorized access or interception. Authentication mechanisms verify user and device identities, preventing unauthorized network access. Additionally, access control policies govern permissions and privileges, curtailing potential attack vectors and enhancing network security [9]. The architecture of satellite wireless communication is depicted in Figure 2 [57].

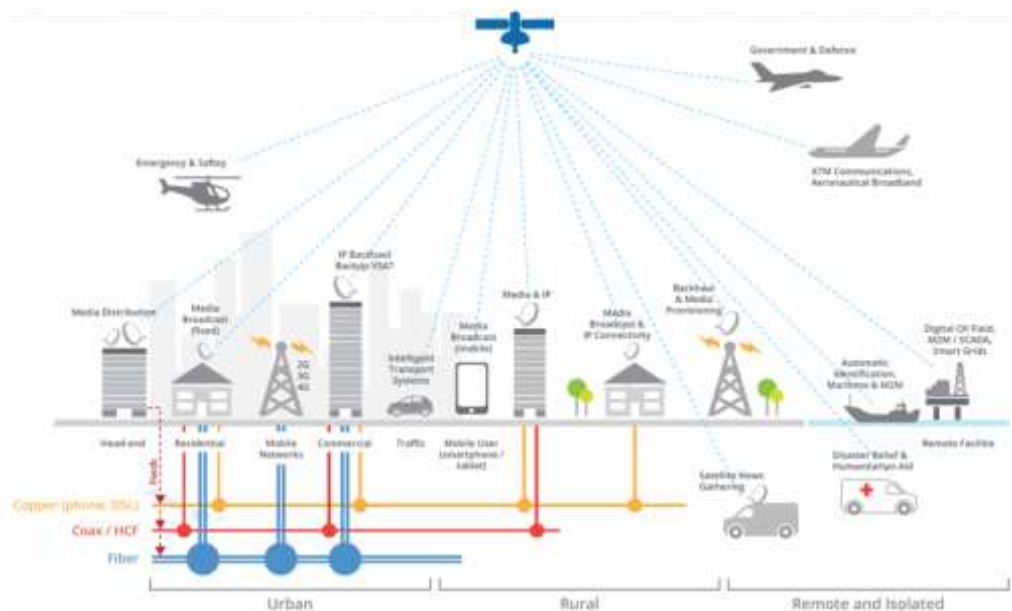

Figure 2: Architecture of Satellite Wireless Communication.

### 3.5. Potential Vulnerabilities:

Despite its advanced architecture and security measures, the Starlink network is susceptible to various vulnerabilities. Cyber threats pose significant risks, including attacks targeting satellites, ground stations, or user terminals. These threats may exploit weaknesses in encryption algorithms, authentication protocols, or software vulnerabilities, compromising data integrity and confidentiality. Furthermore, physical threats such as space debris or natural disasters pose additional risks to network infrastructure, potentially disrupting connectivity [6].

8  Karwan Mustafa Kareem

The interconnected nature of the Starlink network introduces dependencies and interdependencies, amplifying the impact of vulnerabilities or failures. A single point of failure within the satellite constellation, ground infrastructure, or communication links could disrupt connectivity for numerous users, underscoring the importance of redundancy and resilience in the system architecture [7].

In conclusion, the system architecture of Starlink represents a monumental endeavour to provide high-speed internet access globally. By leveraging cutting-edge technology and innovative design, Starlink aims to bridge the digital divide and bring connectivity to underserved regions. However, the architecture also faces challenges, including security vulnerabilities and potential disruptions. To ensure the integrity and availability of the network, robust measures must be implemented to mitigate risks and enhance resilience. Despite these challenges, Starlink stands as a testament to human ingenuity and innovation, offering hope for a more connected future.

## 4. CYBER THREAT LANDSCAPE ANALYSIS

The current cyber threat landscape confronting satellite internet networks, including Starlink, is characterized by a myriad of evolving challenges. Satellites, ground stations, user terminals, and the interconnecting infrastructure are all potential targets for malicious actors seeking to disrupt services, compromise data integrity, or gain unauthorized access. Threat actors may include state-sponsored entities, cybercriminal organizations, or even lone hackers motivated by financial gain, espionage, or ideological reasons.

Satellite internet networks operate within a hostile environment where traditional cybersecurity measures face unique challenges. Unlike terrestrial networks, satellite communications are susceptible to interception, jamming, or spoofing, making them vulnerable to exploitation. Additionally, the sheer complexity and scale of satellite constellations pose challenges for effective threat detection and mitigation.

### 4.1. Identifying Specific Threats and Attack Vectors Targeting Starlink

Several specific threats and attack vectors could potentially target Starlink, posing significant risks to its operations and users:

- **DDoS Attacks:** Distributed Denial of Service (DDoS) attacks can overload Starlink's infrastructure with an overwhelming volume of traffic, causing service disruptions and rendering the network inaccessible to legitimate users [15][13].
- **Spoofing:** Spoofing attacks involve manipulating satellite signals to deceive user terminals or ground stations. Attackers may impersonate legitimate satellites or ground stations, leading to unauthorized access, data manipulation, or interception [16],[14].
- **Signal Interception:** Malicious actors may intercept and eavesdrop on satellite communications, compromising the confidentiality and privacy of transmitted data. Intercepted data could be exploited for espionage, surveillance, or other nefarious purposes [17].
- **Jamming:** Jamming attacks involve transmitting interference signals to disrupt satellite communications. By overpowering legitimate signals, attackers can degrade connectivity, disrupt operations, or render services unusable for targeted regions [18].

- **Cyber Espionage:** State-sponsored actors or cybercriminal groups may engage in cyber espionage activities targeting Starlink to gather intelligence, steal proprietary information, or gain strategic advantages in geopolitical conflicts [19].
- **Cyber Warfare:** In the event of a cyber conflict or military confrontation, satellite internet networks like Starlink could become targets for cyber warfare operations aimed at disrupting critical infrastructure, destabilizing communications, or impairing military capabilities [20].

## 4.2. Discussing the Potential Impact of Cyber Threats on Starlink's Operations and Users

The potential impact of cyber threats on Starlink's operations and users is profound and multifaceted:

- Service Disruptions: DDoS attacks, jamming, or signal interference could disrupt Starlink's operations, leading to service outages and connectivity issues for users. These disruptions may impact critical services, emergency communications, or essential infrastructure relying on satellite connectivity.
- Data Compromise: Spoofing or signal interception attacks could compromise the confidentiality and integrity of data transmitted over Starlink's network. User data, sensitive information, or proprietary data exchanged over the network could be intercepted, manipulated, or stolen by malicious actors.
- Financial Losses: Service disruptions, reputational damage, or regulatory penalties resulting from cyber attacks could lead to significant financial losses for Starlink and its parent company, SpaceX. Moreover, users affected by cyber attacks may incur financial losses due to downtime, data breaches, or compromised services.
- Reputational Damage: High-profile cyber attacks targeting Starlink could tarnish its reputation and erode user trust. Public perception of Starlink's reliability, security, and privacy may be negatively impacted, resulting in customer churn, reduced adoption rates, or diminished market share.
- National Security Implications: Cyber attacks targeting satellite internet networks like Starlink could have far-reaching national security implications. Disruptions to critical infrastructure, emergency communications, or military operations could jeopardize national security interests and compromise strategic capabilities.
- Geopolitical Tensions: Cyber attacks against Starlink could exacerbate geopolitical tensions and trigger diplomatic conflicts between nations. State-sponsored cyber attacks targeting satellite networks may escalate into broader geopolitical confrontations, undermining international stability and cooperation.

In conclusion, the cyber threat landscape facing satellite internet networks like Starlink is dynamic, sophisticated, and ever-evolving. As reliance on satellite connectivity continues to grow, proactive measures to mitigate cyber threats, enhance resilience, and safeguard critical infrastructure are imperative. By adopting robust cybersecurity strategies, collaborating with industry partners, and leveraging advanced technologies, Starlink can mitigate risks, protect its operations and users, and ensure the integrity and availability of its network in an increasingly hostile cyber environment.

## 5. RISK ASSESSMENT FOR STARLINK CYBERSECURITY THREATS

Cybersecurity threats pose significant risks to the operations and users of Starlink, SpaceX's satellite internet service. Conducting a thorough risk assessment is crucial to identify potential threats, evaluate their likelihood and potential impact, and prioritize mitigation efforts. In this analysis, we will delve into the complexities of conducting a risk assessment for Starlink, utilizing methodologies such as the NIST Cybersecurity Framework and OCTAVE Allegro to assess cybersecurity risks comprehensively.

### 5.1. *Understanding Risk Assessment Methodologies*

Risk assessment methodologies provide structured approaches to identify, evaluate, and prioritize cybersecurity risks. Two widely recognized methodologies are the NIST Cybersecurity Framework and OCTAVE Allegro:

#### 5.1.1. *NIST Cybersecurity Framework*

The National Institute of Standards and Technology (NIST) Cybersecurity Framework offers a structured approach to managing cybersecurity risk. It consists of five core functions: Identify, Protect, Detect, Respond, and Recover. These functions provide a framework for organizations to understand, manage, and reduce cybersecurity risks effectively.

The **Identify** function focuses on understanding the organization's cybersecurity risk posture by identifying critical assets, systems, and processes. This includes identifying the assets within the Starlink infrastructure, such as satellites, ground stations, user terminals, communication links, and data centres.

The **Protect** function involves implementing safeguards to protect identified assets from cybersecurity threats. This may include implementing access controls, encryption mechanisms, intrusion detection systems, and security awareness training for employees.

The **Detect** function focuses on identifying cybersecurity events promptly. This involves implementing monitoring and detection capabilities to identify potential threats and vulnerabilities in real-time.

The **Respond** function involves developing and implementing response plans to address cybersecurity incidents effectively. This includes establishing incident response teams, communication protocols, and recovery procedures to minimize the impact of cybersecurity incidents.

The **Recover** function focuses on restoring the organization's capabilities and services following a cybersecurity incident. This includes implementing backup and recovery procedures, conducting post-incident analysis, and implementing lessons learned to improve future incident response [21].

#### 5.1.2. *OCTAVE Allegro*

Operationally Critical Threat, Asset, and Vulnerability Evaluation (OCTAVE) Allegro is a risk assessment methodology developed by the Software Engineering Institute (SEI) at Carnegie Mellon University. OCTAVE Allegro focuses on identifying critical assets, assessing threats and vulnerabilities, and developing risk mitigation strategies tailored to the organization's specific needs.

OCTAVE Allegro begins with defining the organization's mission and goals and identifying its critical assets. For Starlink, critical assets may include its satellite constellation, ground stations, user terminals, and communication infrastructure.

The methodology then assesses threats and vulnerabilities that could impact these critical assets. This includes identifying potential cybersecurity threats such as DDoS attacks, spoofing, signal interception, cyber espionage, and cyber warfare, as well as vulnerabilities arising from insecure communication protocols, outdated software, insufficient encryption mechanisms, or human error.

Once threats and vulnerabilities are identified, OCTAVE Allegro evaluates the likelihood and potential impact of each threat. This allows organizations to prioritize risks based on their severity and likelihood of occurrence.

Finally, OCTAVE Allegro helps organizations develop risk mitigation strategies tailored to their specific needs and risk tolerance. This may include implementing technical controls, enhancing security policies and procedures, and providing employee training and awareness programs [22].

## 5.2. Conducting Risk Assessment for Starlink

Conducting a risk assessment for Starlink involves several steps, including identifying critical assets, identifying threats and vulnerabilities, assessing likelihood and impact, and prioritizing risks.

### 5.2.1. Step 1: Identify Critical Assets

The first step in the risk assessment process is to identify critical assets within the Starlink infrastructure. These assets may include satellites, ground stations, user terminals, communication links, and data centres. Understanding the importance of these assets is essential for prioritizing risk management efforts.

Identifying critical assets involves understanding their role in the operation of the Starlink network and their potential impact on the organization if compromised. For example, satellites are critical assets as they form the backbone of the Starlink network and enable communication between ground stations and user terminals.

### 5.2.2. Step 2: Identify Threats and Vulnerabilities

Once critical assets are identified, the next step is to identify potential cybersecurity threats and vulnerabilities that could impact these assets. Threats may include DDoS attacks, spoofing, signal interception, cyber espionage, and cyber warfare. Vulnerabilities could arise from insecure communication protocols, outdated software, insufficient encryption mechanisms, or human error.

Identifying threats and vulnerabilities requires a comprehensive understanding of the Starlink network architecture, including its components, protocols, and interfaces. This may involve conducting interviews with subject matter experts, reviewing technical documentation, and analysing historical data on cybersecurity incidents.

*5.2.3.   Step 3: Assess Likelihood and Impact*

Using the NIST Cybersecurity Framework or OCTAVE Allegro, the likelihood and potential impact of each identified threat are assessed. Likelihood refers to the probability of a threat occurring, while impact assesses the severity of the consequences if the threat materializes.

Assessing likelihood and impact involves analyzing historical data, threat intelligence, and expert judgment to estimate the probability of a threat occurring and its potential impact on Starlink operations and users. This allows organizations to prioritize risks based on their severity and likelihood of occurrence.

*5.2.4.   Step 4: Prioritize Risks:*

Based on the results of the risk assessment, risks are prioritized by considering their likelihood and potential impact. Risks with high likelihood and high impact are deemed the most critical and require immediate attention. Conversely, risks with low likelihood and low impact may be deemed acceptable or mitigated through standard security measures.

Prioritizing risks involves weighing the potential benefits of risk mitigation efforts against their costs and resource requirements. This may involve conducting cost-benefit analyses, stakeholder consultations, and risk tolerance assessments to determine the appropriate risk management strategies.

In conclusion, conducting a risk assessment of cybersecurity threats to Starlink is essential for identifying, prioritizing, and mitigating risks to its operations and users. By leveraging methodologies such as the NIST Cybersecurity Framework or OCTAVE Allegro, Starlink can systematically evaluate cybersecurity risks, establish risk management priorities, and implement effective mitigation strategies. By prioritizing risk management efforts and proactively addressing cybersecurity threats, Starlink can enhance the resilience and security of its satellite internet service, ensuring the integrity and availability of its network for users worldwide.

## 6.   MITIGATION STRATEGIES

Cybersecurity risks pose significant challenges to the integrity and availability of satellite internet networks, including Starlink, SpaceX's ambitious project. Mitigation strategies play a crucial role in addressing these risks and ensuring the resilience of satellite internet infrastructure. In this analysis, we propose a comprehensive set of mitigation strategies encompassing technical solutions, policy measures, and regulatory frameworks to enhance cybersecurity resilience in satellite internet networks.

### *6.1.   Technical Mitigation Strategies*

*6.1.1.   Encryption Mechanisms*

Implementing robust encryption mechanisms is essential to protect data transmitted within satellite internet networks. End-to-end encryption ensures that data is encrypted throughout its transmission, safeguarding it from interception or tampering by malicious actors. Advanced encryption standards such as AES (Advanced Encryption Standard) with strong key management practices can help mitigate the risk of unauthorized access to sensitive information [23].

### 6.1.2. Authentication Mechanisms

Deploying strong authentication mechanisms is crucial to verify the identities of users and devices accessing satellite internet networks. Multi-factor authentication (MFA) enhances security by requiring multiple forms of authentication, such as passwords, biometrics, or token-based authentication. This helps prevent unauthorized access and strengthens the overall security posture of the network [24].

### 6.1.3. Intrusion Detection Systems (IDS)

Implementing intrusion detection systems (IDS) enables real-time monitoring and detection of suspicious activities or unauthorized access attempts within the satellite internet network. IDS can detect anomalies in network traffic, unusual patterns of behaviour, or known attack signatures, allowing security teams to respond promptly to potential security incidents and mitigate risks before they escalate [25].

### 6.1.4. Secure Network Architecture

Designing and implementing a secure network architecture is essential to minimize attack surfaces and vulnerabilities within satellite internet networks. Segmentation of network infrastructure, enforcing least privilege access controls, and implementing network firewalls help prevent unauthorized access and lateral movement by attackers. Additionally, deploying network intrusion prevention systems (IPS) adds an additional layer of defence against malicious activities [26].

### 6.1.5. Continuous Monitoring and Incident Response

Establishing a robust system for continuous monitoring and incident response enables proactive detection and response to cybersecurity threats in real-time. Security operations centres (SOCs) equipped with advanced security analytics tools can monitor network traffic, detect anomalies, and respond to security incidents promptly. Implementing incident response plans and conducting regular security drills ensures that security teams are well-prepared to handle cybersecurity incidents effectively [27].

## 6.2. Policy and Regulatory Measures

### 6.2.1. Cybersecurity Policies and Standards

Developing and enforcing cybersecurity policies and standards tailored to satellite internet networks is essential to ensure consistent security practices across the industry. These policies should outline requirements for encryption, authentication, access control, data protection, and incident response, aligning with industry best practices and regulatory requirements [30].

### 6.2.2. Regulatory Frameworks

Establishing regulatory frameworks specific to satellite internet networks can help ensure compliance with cybersecurity standards and guidelines. Regulatory bodies can mandate security requirements, conduct audits, and enforce penalties for non-compliance, incentivizing satellite internet providers to prioritize cybersecurity and invest in security measures proactively [29].

### 6.2.3. International Collaboration and Information Sharing

Promoting international collaboration and information sharing among satellite internet providers, regulatory bodies, and cybersecurity organizations facilitates collective efforts to address cybersecurity challenges effectively. Sharing threat intelligence, best practices, and lessons learned enhances situational awareness and strengthens defences against evolving cyber threats in satellite internet networks [28].

In conclusion, mitigating cybersecurity risks in satellite internet networks requires a multi-faceted approach encompassing technical solutions, policy measures, and regulatory frameworks. Implementing robust encryption and authentication mechanisms, deploying intrusion detection systems, and ensuring secure network architecture are essential technical measures to protect against cyber threats.

Additionally, developing cybersecurity policies, establishing regulatory frameworks, and promoting international collaboration enhance cybersecurity resilience and promote a culture of security within the satellite internet industry. By implementing these mitigation strategies comprehensively, satellite internet providers can strengthen the security posture of their networks, safeguard sensitive data, and ensure the integrity and availability of satellite internet services for users worldwide.

## 7. MITIGATION STRATEGIES FOR CYBERSECURITY RISKS IN SATELLITE INTERNET NETWORKS

### 7.1. Encryption Mechanisms

Implementing robust encryption mechanisms is paramount to protect data transmitted within satellite internet networks [31]. End-to-end encryption ensures data remains encrypted throughout transmission, thwarting interception or tampering attempts by malicious actors. Advanced encryption standards, such as AES (Advanced Encryption Standard), coupled with stringent key management practices, help mitigate the risk of unauthorized access to sensitive information [32]. Moreover, leveraging quantum-resistant encryption algorithms offers future-proof protection against emerging threats to encrypted data [33].

### 7.2 Authentication Mechanisms

Deploying strong authentication mechanisms is pivotal to verifying the identities of users and devices accessing satellite internet networks [34]. Multi-factor authentication (MFA) enhances security by necessitating multiple forms of authentication, like passwords, biometrics, or token-based authentication [35]. This mitigates the risk of unauthorized access and bolsters the overall security posture of the network. Additionally, implementing device authentication protocols ensures only authorized devices can connect to the network, curtailing the risk of unauthorized access or device spoofing [36].

### 7.3 Intrusion Detection Systems (IDS)

Implementing intrusion detection systems (IDS) enables real-time monitoring and detection of suspicious activities or unauthorized access attempts within satellite internet networks [37]. IDS can discern anomalies in network traffic, unusual behaviour patterns, or known attack signatures, empowering security teams to respond promptly to potential security incidents and mitigate risks before escalation [38]. Furthermore, employing machine learning-based IDS enhances threat detection capabilities by identifying previously unseen threats based on behavioural analysis and anomaly detection [39].

### 7.4 Secure Network Architecture

Designing and implementing a secure network architecture is pivotal to minimizing attack surfaces and vulnerabilities within satellite internet networks [40]. Segmentation of network infrastructure, enforcing least privilege access controls, and deploying network firewalls help prevent unauthorized access and lateral movement by attackers [41]. Moreover, integrating network segmentation with zero-trust architecture ensures stringent access controls and limits the lateral movement of attackers within the network, enhancing overall security [42]. Additionally, leveraging software-defined networking (SDN) facilitates dynamic network segmentation, allowing for rapid response to emerging threats and changing network conditions [43].

### 7.5 Continuous Monitoring and Incident Response

Establishing a robust system for continuous monitoring and incident response enables proactive detection and response to cybersecurity threats in real-time [44]. Security operations centres (SOCs) equipped with advanced security analytics tools can monitor network traffic, detect anomalies, and respond to security incidents promptly [45]. Implementing incident response plans and conducting regular security drills ensures security teams are well-prepared to handle cybersecurity incidents effectively [16]. Additionally, leveraging Security Orchestration, Automation, and Response (SOAR) platforms streamlines incident response workflows, enabling rapid detection, triage, and remediation of security incidents [47].

### 7.6 Employee Training and Awareness

Investing in employee training and awareness programs is pivotal to fostering a cybersecurity-conscious culture within the organization [48]. Educating employees about common cyber threats, phishing scams, and best practices for securing their devices and data mitigates the risk of insider threats and human error [49]. Additionally, conducting regular security awareness training sessions and simulated phishing exercises reinforces security awareness and empowers employees to recognize and report suspicious activities effectively [50].

### 7.7 Regular Security Audits and Vulnerability Assessments

Conducting regular security audits and vulnerability assessments helps identify and remediate potential security weaknesses within satellite internet networks [51]. Penetration testing, vulnerability scanning, and code reviews enable security teams to proactively identify and address vulnerabilities before they can be exploited by malicious actors [52].

Moreover, leveraging automated vulnerability assessment tools facilitates continuous monitoring of network vulnerabilities and ensures timely remediation of security gaps [53].

In conclusion, implementing robust mitigation strategies is essential to mitigate cybersecurity risks in satellite internet networks like Starlink. By prioritizing the implementation of encryption mechanisms, authentication protocols, intrusion detection systems, secure network architecture, continuous monitoring, employee training, and regular security audits, satellite internet providers can enhance the resilience and security of their networks. Additionally, fostering a culture of security awareness and investing in ongoing research and development are key to staying ahead of emerging cyber threats and ensuring the integrity and availability of satellite internet services for users worldwide.

## 8. CASE STUDIES AND ANALYSIS

Cybersecurity incidents in satellite internet networks have underscored the importance of robust defences and proactive risk mitigation strategies. While specific cases involving satellite internet networks like Starlink may be limited due to the relative novelty of such systems, examining analogous incidents in the broader satellite communications industry provides valuable insights into potential threats and vulnerabilities. This analysis will explore relevant case studies, draw lessons learned, and discuss their implications for Starlink's cybersecurity.

### *8.1.   Case Study 1: The 1998 Solar Sunrise Attack*

One notable cyber incident in the satellite communications sector is the Solar Sunrise attack, which occurred in 1998. While not specifically targeting satellite internet networks, this incident serves as a cautionary tale regarding the vulnerability of critical infrastructure to cyber threats.

Solar Sunrise involved a series of coordinated cyber attacks targeting U.S. military and government computer systems, including those supporting satellite communications. The attackers exploited vulnerabilities in unpatched Solaris operating systems to gain unauthorized access to networks, compromising sensitive data and disrupting operations [54].

**Lessons Learned:**
- Vulnerability of Legacy Systems: The Solar Sunrise incident highlighted the risks posed by outdated and unpatched systems. Legacy infrastructure, including satellites and ground stations, may be susceptible to similar attacks if not adequately maintained and updated.
- Need for Vigilance: The incident emphasized the importance of continuous monitoring and threat detection to identify suspicious activities promptly. Early detection can mitigate the impact of cyber attacks and prevent unauthorized access to critical systems.
- Importance of Response Planning: Effective incident response procedures are essential to minimize the impact of cyber attacks and restore operations swiftly. Preparedness includes establishing incident response teams, communication protocols, and recovery strategies tailored to the unique challenges of satellite networks.

**Implications for Starlink:**

While Starlink's advanced architecture may mitigate some risks associated with legacy systems, lessons from the Solar Sunrise attack underscore the importance of proactive cybersecurity measures. Starlink must prioritize regular system updates, robust monitoring capabilities, and comprehensive incident response planning to defend against similar threats.

### 8.2. Case Study 2: GPS Spoofing Incidents

GPS spoofing incidents represent another significant cybersecurity challenge for satellite communications, with implications for satellite internet networks. In a GPS spoofing attack, malicious actors manipulate satellite signals to deceive receivers, leading to inaccurate location data and potential disruption of critical services.

One notable example occurred in 2013 when researchers demonstrated the ability to spoof GPS signals and take control of a yacht's navigation system, altering its course without detection. While this incident did not directly impact satellite internet networks, it underscores the potential consequences of spoofing attacks on satellite-based systems [55].

**Lessons Learned:**
- Critical Infrastructure Vulnerability: GPS spoofing incidents highlight the vulnerability of critical infrastructure, including satellite navigation systems, to manipulation by malicious actors. Satellite internet networks relying on accurate positioning data may face similar risks if GPS signals are compromised.
- Need for Signal Authentication: Implementing robust authentication mechanisms for satellite signals can help verify their authenticity and detect spoofing attempts. Techniques such as signal encryption and digital signatures can enhance the integrity and reliability of transmitted data.
- Importance of Redundancy: Redundant systems and backup mechanisms are essential to mitigate the impact of GPS spoofing incidents and ensure continuity of operations. Diverse communication paths and alternative navigation methods can help maintain connectivity and reliability in the event of an attack.

**Implications for Starlink:**

While Starlink's architecture may reduce reliance on GPS for positioning, the potential for spoofing attacks highlights the importance of signal authentication and redundancy in safeguarding satellite internet networks. Implementing measures to verify signal integrity and establish backup communication channels can enhance the resilience of Starlink's infrastructure against similar threats.

### 8.3. Case Study 3: Intentional Interference

Intentional interference with satellite communications, often referred to as jamming, poses a significant threat to satellite internet networks' reliability and availability. While jamming incidents may vary in scale and sophistication, they can disrupt services, degrade signal quality, and impair connectivity for users.

In 2019, reports emerged of widespread GPS jamming incidents in the Middle East, affecting maritime and aviation navigation systems. While the exact cause of these disruptions remains unclear, intentional interference by state or non-state actors was suspected, highlighting the potential for malicious jamming activities to impact satellite-based services [56].

**Lessons Learned:**
- Impact of External Factors: Satellite internet networks are susceptible to external interference, including jamming and signal disruption. Understanding and mitigating environmental factors, geopolitical tensions, and malicious activities are essential to ensuring uninterrupted service delivery.
- Need for Resilient Infrastructure: Building resilient infrastructure capable of withstanding and recovering from interference events is crucial for maintaining service continuity. Redundant systems, alternative communication paths, and rapid response mechanisms can help mitigate the impact of jamming incidents.
- Collaborative Response: Addressing intentional interference requires collaboration among satellite operators, regulatory authorities, and international stakeholders. Sharing threat intelligence, coordinating response efforts, and implementing countermeasures collectively can enhance the industry's ability to combat malicious activities effectively.

**Implications for Starlink:**

While Starlink's extensive satellite constellation may mitigate some risks of signal interference, intentional jamming remains a potential threat to its operations. Implementing robust jamming detection mechanisms, enhancing signal resilience, and fostering collaboration with relevant stakeholders can strengthen Starlink's defences against intentional interference incidents.

Cyber incidents in satellite communications, while diverse in nature, provide valuable insights into potential threats and vulnerabilities facing satellite internet networks like Starlink. By analysing case studies such as the Solar Sunrise attack, GPS spoofing incidents, and intentional interference events, valuable lessons can be learned regarding the importance of proactive cybersecurity measures, vulnerability management, and collaborative response efforts. As Starlink continues to expand its global internet infrastructure, incorporating these lessons into its cybersecurity strategy will be essential to safeguarding its operations and ensuring the integrity and availability of its services for users worldwide.

## 9. METHODOLOGY

This research employed a systematic methodology to comprehensively investigate the cybersecurity challenges and mitigation strategies pertinent to satellite internet networks, with a specific emphasis on Starlink. The methodology encompassed several key stages, beginning with an extensive literature review to establish a foundational understanding of the subject matter. Academic databases such as IEEE Xplore, PubMed, and Google Scholar were queried using relevant keywords, including "satellite internet," "cybersecurity," "Starlink," and "mitigation strategies." The retrieved scholarly articles, research papers, industry reports, and official documentation served as primary sources of data for the study. Additionally, secondary sources such as news articles and white papers were consulted to gather diverse perspectives on the topic.

Following data collection, a rigorous analysis phase was undertaken to identify key themes, trends, and insights within the literature. Data analysis techniques, including thematic analysis and content analysis, were employed to categorize and interpret the information effectively. Comparative analysis was conducted to evaluate different approaches proposed in the literature and identify gaps in current research. The synthesized findings from the literature review and data analysis formed the basis for addressing the research objectives and questions posed in the study.

Throughout the research process, ethical considerations were carefully observed to uphold academic integrity and respect for intellectual property rights. Proper citation and acknowledgment of sources were maintained, and ethical guidelines for conducting research were adhered to. It is important to note certain limitations inherent in the methodology, including reliance on secondary data sources and the dynamic nature of cybersecurity threats. However, efforts were made to mitigate these limitations by consulting diverse and reputable sources and focusing on recent publications.

In conclusion, the systematic methodology employed in this research provided a structured approach to explore the cybersecurity landscape of satellite internet networks, particularly focusing on Starlink. Adjustments were made based on the specific requirements and scope of the study, ensuring robustness and reliability in the research process.

## 10. CONCLUSION

The research conducted has illuminated the cybersecurity challenges confronting satellite internet networks, with a specific focus on Starlink. Through an exhaustive analysis encompassing the threat landscape, system architecture, risk assessment, mitigation strategies, and case studies, several key findings have surfaced. These findings carry significant implications for the cybersecurity of Starlink and other satellite internet networks, while also suggesting avenues for future research to tackle emerging challenges.

Satellite internet networks encounter a diverse array of cybersecurity threats, ranging from DDoS attacks to signal interception, posing risks to communication system reliability and integrity. The implementation of robust encryption, authentication, intrusion detection, and incident response mechanisms emerge as crucial in mitigating these risks. Technical, policy, and regulatory measures are deemed essential to bolster the resilience of satellite infrastructure against evolving cyber threats. Moreover, analysis of historical cyber incidents provides valuable insights into potential vulnerabilities and mitigation strategies for satellite internet networks.

The study's findings hold significant implications for Starlink and other satellite internet networks. Starlink must prioritize the implementation of robust cybersecurity measures to safeguard its infrastructure, users, and data. This includes proactive monitoring, threat detection, incident response planning, and collaboration with industry partners. Compliance with cybersecurity regulations and industry standards is imperative for ensuring the security and resilience of satellite internet networks. Starlink should adhere to relevant regulatory frameworks and engage with regulatory authorities to address compliance requirements effectively. Additionally, continued investment in research and development is necessary to address emerging cybersecurity challenges and technological advancements in satellite internet networks.

Looking ahead, several future research directions emerge to further enhance the cybersecurity resilience of satellite internet networks. Exploring security automation through artificial intelligence, machine learning, and automation technologies can bolster defences and response capabilities. Investigating the adoption of quantum-safe cryptographic algorithms can mitigate the risk of quantum computing attacks on satellite communication systems. Conducting comprehensive resilience testing and simulations to assess network robustness against cyber threats and disruptive events is essential. Furthermore, contributing to the development of international cybersecurity policies and standards specific to satellite internet networks can foster collaboration and information sharing among stakeholders.

In conclusion, addressing cybersecurity challenges in satellite internet networks necessitates a holistic approach encompassing technical innovation, regulatory compliance, collaboration, and ongoing research. By proactively implementing measures and investing in cybersecurity resilience, Starlink and other satellite internet providers can uphold the integrity, availability, and confidentiality of their services in an interconnected and digital era.

## 11. ACKNOWLEDGEMENTS

We would like to extend our sincere gratitude to all those who contributed to the completion of this research project. Firstly, we express our appreciation to the participants who generously shared their time and knowledge, without whom this research would not have been possible. Their contributions provided valuable perspectives and enriched the depth of our analysis.

We also acknowledge the support and resources provided by Computer Science Dep., College of Basic Education/ University of Sulaimani. The facilities and access to literature greatly facilitated our research endeavours.

Furthermore, we express our gratitude to our friends and family for their unwavering encouragement and understanding during the course of this project. Their patience and encouragement kept us motivated during challenging times.

## Biographies

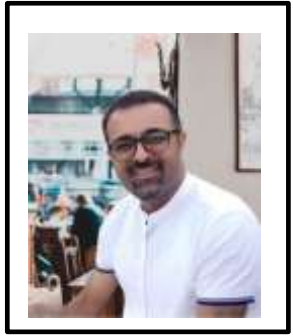

**Karwan Mustafa Kareem** is an accomplished authority in the fields of IT, Cyber Security, and Cyber Crimes. Born in Halabja in 1983, he completed his primary and secondary education in the same city. Kareem earned his Bachelor's degree in Computer and Mathematics from Sulaimani University in 2007. Later, he pursued a Master's degree in Advanced Computer Science from the University of Huddersfield, Department of Computing and Engineering in 2011. With a wealth of experience spanning over 20 years in cyber security and cybercrimes, Kareem presently serves as a tutor at Sulaimani University and also serves as the CEO of Diamond Company for Cybersecurity Services.